\documentclass{amsart}

\usepackage{amsmath}

\usepackage{mathrsfs}
\usepackage{amssymb}

\usepackage{graphicx}

\begin{document}

\title[Modular Theory and Quantum Entanglement]{
Modular Theory for Operator Algebra\\
 in Bounded Region of Space-Time\\
and  Quantum Entanglement}

\author{Daisuke Ida}

\author{Takahiro Okamoto}

\author{Miyuki Saito}
\address{Department of Physics, Gakushuin University, Tokyo 171-8588, Japan.}

\date{Mon., May 27th, 2013}

\begin{abstract}
We consider the quantum state seen by an observer in the diamond-shaped region,
which is a globally hyperbolic open submanifold of the Minkowski space-time.
It is known from the operator-algebraic argument
that the vacuum state of the quantum field transforming covariantly under the conformal group
looks like a thermal state on the von Neumann algebra generated by the field operators
on the diamond-shaped region of the Minkowski space-time.
Here, we find,
in the case of the free massless Hermitian scalar field in the 
2-dimensional Minkowski space-time,
 that such a state can in fact be identified with a certain
entangled quantum state.
By doing this, we obtain the thermodynamic quantities such as the Casimir energy
and the von Neumann entropy of the thermal state in the diamond-shaped region,
and show that the Bekenstein bound for the entropy-to-energy ratio is saturated.
We further speculate on a possible information-theoretic 
interpretation of the entropy in terms of the probability density functions
naturally determined from the Tomita-Takesaki modular flow 
in the diamond-shaped region.
\end{abstract}

\maketitle

\section{Background and Motivation}
We often regard the quantum state of a field on the space-time
as being a pure state that has the zero von Neumann entropy.
Of course, this does not imply that an observer always has a
perfect knowledge of the quantum field.
Rather, each observer would not be able to distinguish it
from a certain mixed state, and the identified mixed state would in general
depend on the observer's trajectory and the measuring means available. 
Thus, each observer has his own nonzero 
von Neumann entropy for the quantum states of the field.

For example, let us consider an observer with a finite lifetime
whose world-line is a 
 timelike segment 
in the space-time
bounded by future and past end points, 
and the measurements of the quantum field by him in terms of 
an apparatus located at each space-time point.
When this observer sends a command 
to a remote measuring apparatus,
the apparatus immediately performs a measurement of the quantum field
on the corresponding space-time point 
and the result is returned to the observer. 
If this is the only way for the observer to measure the quantum state of 
the field, the set of points from which the observer can get the information
is the intersection of the chronological future 
and the chronological past of the  observer's world-line,
which we call,  for obvious reasons,
 the ``diamond region''
associated with the observer.
The limitation of the observed region would
cause the loss of the information on the quantum state of the
field.
This can be heuristically understood from  general considerations as follows.

In general, a quantum measurement can be reduced to the evaluation of
the expectation value of a non-negative self-adjoint operator belonging to a
$C^*$-algebra $\mathscr{A}$.  
In the quantum field theory, 
the corresponding $C^*$-algebra $\mathscr{A}$ might be regarded as 
the von Neumann algebra $\mathscr{A}(M)$ constructed from the field operators
on the space-time $M$.
(Though the polynomial $*$-algebra generated by field operators is not 
a von Neumann algebra, for the field operators are unbounded,
one can define a von Neumann algebra $\mathscr{A}(O)$ constructed from 
field operators on the open submanifold $O$ of $M$, 
if $O$ is $M$ itself, a diamond region, 
a so-called Rindler wedge, or their image under a Poincar\'e transformation~\cite{DSW86,Ha96}.
More precisely, the von Neumann algebra $\mathscr{A}(O)$ is the double commutant
of the $C^*$-algebra generated by the projection operators composing the field operators
smeared by test functions
with  support in $O$.)

However, not all the projection operators in $\mathscr{A}(M)$ are 
available for every observer.
Rather, the available projection operators, or more generally 
non-negative self-adjoint operators, generate a proper von Neumann subalgebra
of $\mathscr{A}(M)$, which would be regarded as the algebra of physical 
quantities for the observer. 
For  an observer with a finite lifetime, 
the corresponding von Neumann subalgebra of physical quantities
would be $\mathscr{A}(O)$, where
$O$ is the diamond region associated with the observer.

On the other hand, a quantum state $\omega:\mathscr{A}\to \boldsymbol{C}$ 
on a $C^*$-algebra $\mathscr{A}$
is 
a pure state if and only if
the GNS representation of $\mathscr{A}$ associated with the quantum state
$\omega$ is irreducible.
However, the GNS representation
of its $C^*$-subalgebra $\mathscr{A}'$ associated with the restriction 
of $\omega$ to $\mathscr{A}'$ is not always irreducible.
If it is reducible, the quantum state $\omega$ is indistinguishable from a
certain
mixed state in terms of any quantum measurements solely of the operators
in $\mathscr{A}'$.

Hence, an observer with a finite lifetime would perceive a certain mixed
state.
Then, how does the vacuum state in the Minkowski space-time look like
for the observer with the finite lifetime?

In the case of the conformally invariant Hermitian scalar field,
Martinetti and Rovelli~\cite{MR03} conclude that such an observer will see a
certain thermal state.
Their reasoning is based on the conformal invariance of the vacuum state and 
the conformal equivalence between the diamond region and the Rindler wedge.
The outline of their argument is as follows.

Let $W$ be the Rindler wedge, which is the open submanifold of 
the $n$-dimensional
Minkowski space-time $(n\ge 2)$ specified by $x^1> |x^0|$ in terms of
the standard time coordinate $x^0$ and one of the standard spatial 
coordinates $x^1$
in the Minkowski space-time.
The Rindler wedge $W$ is globally static in the sense that
the Lorentz boost generated by the Killing vector field 
$x^1\partial_0+x^0\partial_1$ 
acts isometrically on $W$.
A uniformly accelerated observer following an orbit of the Lorentz boost
in the Poincar\'e invariant vacuum state 
would find himself apparently in a thermal bath with the temperature 
proportional to the
proper acceleration. This is well known as the Unruh effect~\cite{Unruh}.

One of rigorous explanations of the Unruh effect is given by 
the Bisognano-Wichmann  theorem~\cite{BW75}.
This theorem shows that
the von Neumann algebra $\mathscr{A}(W)$ gives in an essential way 
an example of the application of
the Tomita-Takesaki modular theory~\cite{Tak70} of operator algebras.
According to the Tomita's fundamental theorem in the modular theory,
given a von Neumann algebra $\mathscr{A}$ acting on a Hilbert space $H$,
and a cyclic and separating vector $|\Omega\rangle\in H$,
there uniquely exists the one-parameter group of automorphism $\{\sigma_s\}$
acting on $\mathscr{A}$, which is called the modular flow.
Furthermore, the modular flow is subject to the Kubo-Martin-Schwinger (KMS) condition
with respect to the vector state corresponding to $|\Omega\rangle$, which
means that $|\Omega\rangle$ is identified with a thermal state.
The Bisognano-Wichmann theorem states that in the case of $\mathscr{A}=\mathscr{A}(W)$,
 $|\Omega\rangle$ corresponds to the Poincar\'e invariant vacuum, and hence
the vacuum is subject to the KMS condition, where the generator of the Lorentz boost plays a
 role of the Hamiltonian.
Thus, the modular flow here can be seen as the
geometric flow generating the time translation in $W$.

A relativistic quantum field in the Minkowski space-time is 
often assumed to transform covariantly under the Poincar\'e group~\cite{SW64}.
If we further require the covariance under the conformal group,
and the conformal invariance of the vacuum state,
we can, in a sense, map the geometric modular flow in the Rindler wedge
$W$ to that in the conformal image of $W$.
(Though in the case of $n=2$, there is no vacuum state invariant under 
the conformal group, it is sufficient to consider a state invariant
under the projective conformal group, which is 
the subgroup generated by the dilatations, the
special conformal transformations
and the Poincar\'e transformations.)

In fact, Hislop and Longo show that 
for the quantum field in the diamond region $O$,
the conformally invariant vacuum is subject to the KMS condition~\cite{HL82},
which relies on the conformal equivalence between the Rindler wedge 
and the diamond region.

Martinetti and Rovelli interpret the modular flow 
as determining the ``thermal time'' in $O$,
and this leads to the notion of the ``diamond temperature''
which is the proper temperature for the observer following 
the modular flow~\cite{MR03}.
The relevant observer in $O$ is the inertial observer or 
the uniformly accelerated observer with the finite lifetime.
A remarkable point here is that even an inertial observer may
perceive a nonzero temperature.
Another feature of the diamond temperature is that it in general diverges
around the future and past end points of the observer's world-line.

It is not clear whether the behavior of the diamond temperature as above 
is universal one or whether it 
is peculiar to the operator-algebraic method.
Hence, we would like to verify the diamond temperature
in terms of 
the standard method~\cite{BD84} via the
determination of the Bogoliubov transformation between different Fock representations.
We will see that 
it gives 
 the same temperature
as that derived by Martinetti and Rovelli.
Then, we discuss the thermodynamic quantities such as the Casimir energy and the
quantum entanglement entropy for the observer with a finite lifetime
based on the standard quantum field theory.
We further introduce the probability density function
 naturally determined by the modular flow
in the diamond region, and attempt to give the information-theoretic interpretation
of the entropy of the diamond region.

\begin{figure}[htbp]
\centering
\includegraphics[width=.4\linewidth]{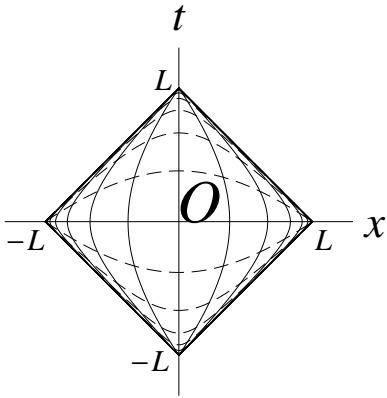}
\caption{The diamond region $O$ of the Minkowski space-time is
  depicted.
The solid curve denotes constant $X$ and
the dashed curve represents constant $T$ [see Eq.~(\ref{eq:TX})]. }
\label{fig:Diamond}
\end{figure}

In this paper, we consider the free massless Hermitian scalar field in the $2$-dimensional
Minkowski space-time $M$, which transforms covariantly under the
projective conformal group. 
We use the natural unit system in which $c=\hbar=1$.
The diamond region $O$ is specified by $|t|+|x|<L$ with
a length parameter $L$, when the Lorentzian metric
is written as  
$ds^2=-dt^2+dx^2$ (Fig.~\ref{fig:Diamond}).

\section{
Thermal State in Diamond Region
}
The modular flow in $O$ coincides with the geometric flow generated by
the conformal Killing vector field, which is timelike in $O$.
This conformal Killing vector field naturally defines  
the positive frequency modes of the Hermitian scalar field for observers
following the modular flow.
In fact, we define the positive frequency modes as the
conformal image of the positive frequency solutions defined on the Rindler wedge $W$,
under the conformal diffeomorphism: $W\to O$, 
which pushes forward the timelike Killing vector field in $W$
 to the conformal Killing vector field in $O$.
More precisely, if the Lorentzian metric $g^O_{\mu\nu}$ in $O$ 
is conformally equivalent to the Lorentzian metric $g^W_{\mu\nu}$ in $W$
as  $g^O_{\mu\nu}=C^2g^W_{\mu\nu}$, and $\xi^\mu$ 
is the timelike Killing vector field with respect to $g^W_{\mu\nu}$,
then $\xi^\mu$ is the conformal Killing vector field with respect to $g^O_{\mu\nu}$.
The positive frequency mode $\chi^{O}_\omega$ in $O$ is 
required to satisfy the eigenvalue equation on $O$
\begin{align*}
\xi^\mu \partial_\mu \chi^{O}_\omega=
-i\omega \chi^{O}_\omega
\end{align*}
for $\omega>0$.

 The null coordinates $U^{\pm}\in {\boldsymbol R}^1$ covering $O$ are defined by 
\begin{align*}
  u^{\pm}=L\tanh (U^{\pm}/L),
\end{align*}
where $u^{\pm} =t\pm x$ are the Minkowski null coordinates.
Then, the positive frequency modes in $O$ are subject to
\begin{align*}
\left({\partial\over \partial {U^+}}+{\partial\over\partial{U^-}}\right)\chi^{O}_\omega=-i\omega\chi^{O}_\omega,\ {\partial^2\over \partial {U^+}\partial{U^-}}\chi^{O}_\omega=0.
\end{align*}
Therefore, the positive frequency mode in $O$ consists of 
\begin{align*}
\chi_\omega^{O\pm} ={1\over \sqrt{4\pi\omega}}\exp({-i\omega U^{\pm} }).
\end{align*}
For the later convenience, we introduce the mode functions
\begin{align*}
  \chi_\omega^{\pm} ={1\over \sqrt{4\pi\omega}}\exp({-i\omega U^{\pm}(u^\pm)})\theta(L-|u^\pm|)
\end{align*}
as extension of $\chi_\omega^{O\pm}$ to $M$,
where we call $\chi_\omega^+$ the ingoing mode, and $\chi_\omega^-$ the outgoing mode,
and 
these mode functions are normalized with respect to the Klein-Gordon inner product.

 On the other hand, by continuing analytically the positive frequency modes $\chi_\omega^{O\pm}$
 to $M$, we obtain
\begin{align}
\widetilde{\chi}_\omega^{\pm}&={{N_\omega}\over \sqrt{4\pi\omega}}\Bigl({L+{u^{\pm}}\over L-{u^{\pm}}}\Bigl)^{-{i}L\omega/2}
\nonumber
\\
&={{N_\omega}\over \sqrt{4\pi\omega}}\times\left\{ \begin{array}{ll} 
\exp({-i \omega U^{\pm}}),& \mbox{for $|u^{\pm}|<L$} \\
e^{-{\pi L}\omega/2}\exp({-i \omega U^{\pm}_{\rm ex}}), & \mbox{for $|u^{\pm}|>L$}
\end{array} \right.\label{inmode}\\
N_{\omega}&=(1-e^{-\pi L \omega })^{-1/2},\nonumber
\end{align} 
where the null coordinates $U_{\rm ex}^\pm$ are defined by
\begin{align*}
u^{\pm}=L\coth\frac{U_{\rm ex}^{\pm}}{L}
\end{align*}
for the regions: $|u^{\pm}|>L$.
Although there are two options to extend $\chi_\omega^{O\pm}$ to $|u^{\pm}|>L$ 
corresponding to the double signs in the relation $\log(-1)=\pm i\pi$,
we remove this ambiguity by requiring that
 $\widetilde\chi_\omega^{\pm}$ correspond to the positive frequency modes with respect
to the Poincar\'e invariant vacuum.

 The positive frequency modes complement to
$\{\chi_\omega^{O\pm}\}$ are determined as
\begin{align*}
\chi_\omega^{{\rm ex}\pm}={1\over \sqrt{4\pi\omega}}\exp({i\omega U_{\rm ex}^{{\pm}}})
\theta(|u^\pm|-L),
\end{align*} 
where we set the sign in the exponent to positive
for $U^{\pm}_{\rm ex}$ are past-directed.
The analytic extension of $\chi_\omega^{{\rm ex}\pm}$ 
from the regions: $|u^\pm|> L$
to $M$ is obtained in the form
\begin{align}
\label{exmode}
\widetilde{\chi}_{\omega}^{{\rm ex}\pm}={{N_\omega}\over \sqrt{4\pi\omega}}
\times\left\{ \begin{array}{ll} 
e^{-{\pi L}\omega/2}\exp({i \omega U^{{\pm}}}), &\mbox{for $|u^{{\pm}}|<L$} \\
\exp({i \omega U^{{\pm}}_{\rm ex}}), & \mbox{for $|u^{{\pm}}|>L$}.
\end{array} \right.
\end{align}

%
\begin{figure}[htbp]
\centering
  \includegraphics[width=1.0\linewidth ]{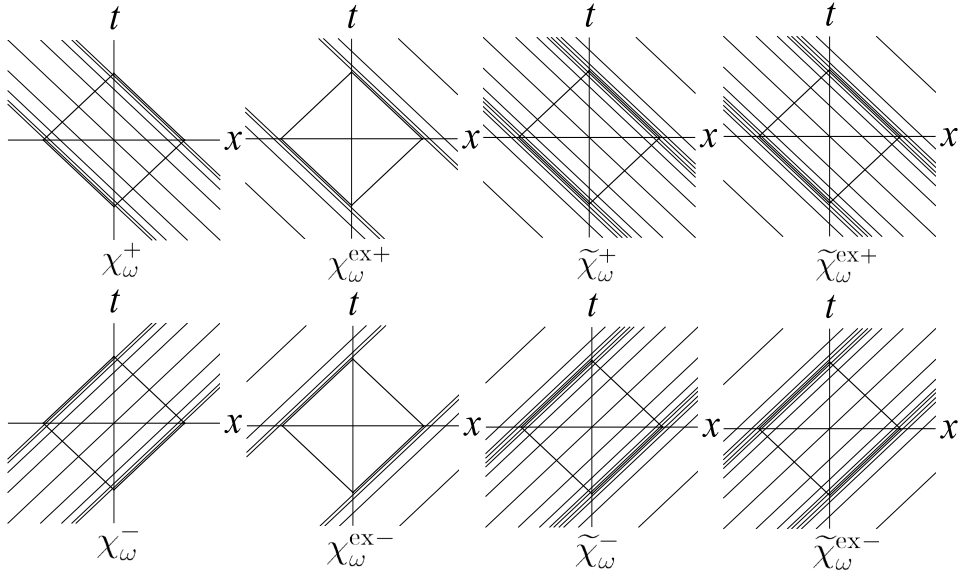}
 \caption{
For two sets of mode functions 
$(\chi_\omega^\pm, \chi_\omega^{\rm ex\pm})$ and $(\widetilde\chi_\omega^\pm,\widetilde\chi_\omega^{\rm ex\pm})$,
the constant phase lines are schematically depicted. 
}
 \label{fig:modefunctions}
\end{figure}

Now, let us derive the Bogoliubov transformation
between the Poincar\'e invariant vacuum and the vacuum defined by the conformal
time flow in the diamond region $O$. 

We can expand the field operator in the form
\begin{align*}
  \phi=&\int_0^\infty d\omega (
{b}^{+}_{\omega}\chi_\omega^+
+{b}^{-}_{\omega}\chi_\omega^-\\
&+{b}^{{\rm ex}+}_{\omega}\chi_\omega^{{\rm ex}+}
+{b}^{{\rm ex}-}_{\omega}\chi_\omega^{{\rm ex}-}
+{\rm H.c.}).
\end{align*}
Then,  the vacuum state in the diamond region $|0;O\rangle$ is defined in terms of 
the annihilation operators $({b}^{\pm}_{\omega},{b}^{{\rm ex}\pm}_{\omega})$
as
\begin{eqnarray*}
{b}^{\pm}_{\omega}|0;O \rangle={b}^{{\rm ex}\pm}_{\omega}|0;O\rangle=0.
\end{eqnarray*}

On the other hand, the Poincar\'e invariant vacuum is defined by 
the set of modes $(\widetilde\chi_\omega^\pm,\widetilde\chi_\omega^{{\rm ex}\pm})$.
More precisely, by writing the mode expansion of the field operator as
\begin{align*}
  \phi&=\int_0^\infty d\omega
(a_\omega^+\widetilde\chi_\omega^++a_\omega^-\widetilde\chi_\omega^-\\
&+a_\omega^{{\rm ex}+}\widetilde\chi_\omega^{{\rm ex}+}
+a_\omega^{{\rm ex}-}\widetilde\chi_\omega^{{\rm ex}-}+{\rm H.c.}),
\end{align*}
the Poincar\'e invariant vacuum $|0;M\rangle$ is defined by
\begin{align*}
{a}^{\pm}_{\omega}|0;M \rangle={a}^{{\rm ex}\pm}_{\omega}|0;M\rangle=0.
\end{align*}
From Eqs.~(\ref{inmode}) and (\ref{exmode}),
the transformation between the two sets of mode functions 
turns out to be
\begin{align*}
\widetilde{\chi}_{\omega}^{\pm}&
=N_{\omega}\left(\chi_{\omega}^{\pm}+e^{-{\pi L}\omega/2}(\chi^{{\rm ex}\pm}_{\omega})^*  \right),\\
\widetilde{\chi}_{\omega}^{{\rm ex}\pm}&
=N_{\omega}\left( \chi_{\omega}^{{\rm ex}\pm}+e^{-{\pi L}\omega/2}({\chi}_{\omega}^{\pm})^* \right).
\end{align*}
This leads to the 
Bogoliubov transformation of the  creation and annihilation operators as
\begin{align*}
{a}_{\omega}^{\pm}&=N_{\omega}\left({b}^{\pm}_{\omega}-e^{-{\pi L}\omega/2}{b}^{{\rm ex}\pm\dagger}_{\omega} \right),\\
{a}_{\omega}^{{\rm ex}\pm}&=N_{\omega}\left( 
{b}^{{\rm ex}\pm}_{\omega} -e^{-{\pi L}\omega/2}{b}_{\omega}^{\pm\dagger}
\right).
\end{align*}
From this, we see that the vacuum $|0;M\rangle$ is also written 
formally as
\begin{align*}
|0;M\rangle&=Z^{-1/2}\\
&\times\prod_{\omega}\exp[{e^{-\pi L\omega /2}
({b}_{\omega}^{+\dagger}{b}_{\omega}^{{\rm ex}+\dagger} 
+ {b}_{\omega}^{-\dagger}{b}_{\omega}^{{\rm ex}-\dagger})}]
|0;O\rangle,\\
Z&=\prod_{\omega }(1-e^{-\pi L \omega})^{-2}.
\end{align*}
By taking the partial trace of the density operator over the subsystem
generated by the operators $b_\omega^{{\rm ex}\pm\dag}$,
we obtain the Gibbs state with the inverse temperature $\beta=\pi L$ as
\begin{align*}
  \rho^O&=Z^{-1}e^{-\pi L H^O},\\
H^O&=\int_0^\infty d\omega~\omega(b_\omega^{+\dag} b_\omega^++b_{\omega}^{-\dag} b_{\omega}^-).
\end{align*}

 It should be noted, however, that there is an ambiguity in the normalization
of the conformal Killing vector field, $\xi^\mu\mapsto \alpha\xi^\mu$,
which affects the inverse temperature
as $\beta\mapsto \alpha^{-1}\beta$.
The invariant inverse temperature is given by
$\beta^O=\beta\sqrt{-\xi_\mu\xi^\mu}$,
which is regarded as the local inverse temperature 
associated with the observer following the 
flow determined by the conformal Killing vector.
In terms of the proper time $\tau$ of the observer,
it becomes
\begin{align*}
\beta^O&={2\pi\over La^2}(\sqrt{1+a^2 L^2}-\cosh(a\tau)),~~
\tau\in (-\tau_a,\tau_a),\\
\tau _a&=a^{-1}{\rm arcsinh}( aL),
\end{align*}
where $a$ denotes the proper acceleration of the observer and 
$2\tau_a$ is the proper length of his lifetime.
This is identical with the diamond temperature of  Martinetti and
Rovelli~\cite{MR03}.
For we have explicitly determined the quantum state in the diamond region,
we would be able to verify this peculiar behavior of the proper temperature 
along the observer with a finite lifetime by constructing a 
concrete model of the particle detector.

Note that this result does not 
immediately imply that
a real thermometer with a finite lifetime in the Minkowski space-time
indicates a nonzero temperature.
In principle, if a thermometer is prepared such that it interacts
only with the field modes in the diamond region, it would indicate 
a finite temperature. 
However, such a preparation would be extremely  difficult, that is
an arbitrarily prepared thermometer would in general be inevitably coupled 
with
field modes outside the diamond region.

Thus, we come to the same conclusion
with different independent arguments, 
which is the evidence that
the diamond temperature has the universal significance.
We also note that the tunneling approach recently proposed by 
Banerjee and Majhi~\cite{PW00,BM09},
which is another independent method to obtain the temperature of the subsystem,
also gives the same temperature,
though we don't state details here.

 To promote a better understanding of the thermodynamics of an observer 
in $O$, we try to find the expression for the energy and the entropy
in the diamond region. 
Firstly, we compute the expectation value of the stress-energy operator $T^O_{\mu\nu}$ 
for the observer in $O$
with respect to the Poincar\'e invariant vacuum. 
The operator $T^O_{\mu\nu}$ is defined in $O$ by
\begin{align*}
{T}^O_{\mu\nu}=:{\phi}_{,\mu}{\phi}_{,\nu}:-\frac{1}{2}g^O_{\mu \nu}:{\phi}^{,\alpha}{\phi}_{,\alpha} :,
\end{align*}
where the colons denote  normal ordering with respect to 
the vacuum $|0;O\rangle$,
and the metric $g^O_{\mu\nu}$ in $O$ is written as 
\begin{align}
g^O& = 
{ -dT^2 +dX^2 \over
\cosh^2 (U^{+}/L)\cosh^2 (U^{-}/L)},\nonumber\\
T&=\frac{U^{+}+U^{-}}{2},~~~~X=\frac{U^{+}-U^{-}}{2}.
\label{eq:TX}
\end{align}

Noting that the Hamiltonian $H^O$ defining the present thermal state
can be written as the spatial integral of $T^O_{TT}$, 
we formally obtain its expectation value as
\begin{align*}
E^O:=&  \langle 0;M| H^O |0;M\rangle
=\int_{-\infty}^\infty dX\langle 0;M| T^O_{TT} |0;M\rangle\\
=&\delta(0)\int_0^\infty d\omega{2\omega\over e^{\pi L\omega}-1}={\delta(0)\over 3L^2}.
\end{align*}
Here it should not be interpreted this divergent result as that the infinite
energy has been confined in a bounded region. 
This quantity is merely the expectation value of the ideal Hamiltonian operator
such that the vacuum state projected onto the diamond region becomes
the thermal state with respect to it, and hence 
it does not imply the instability of the Minkowski space-time.
The divergent factor $\delta(0)$ here 
is controlled by introducing the cutoff scale $\ell$
in the following manner.
Let the scalar field $\phi$ be confined within the interval $-\ell< X<\ell$,
and let $T^O_{\mu\nu}(\ell)$ be the corresponding stress-energy operator.
Then, the frequency of the scalar field is discretized as 
$\omega_n=n\pi/\ell$ ($n=1,2,\cdots$), and  the energy is regarded as the limit
\begin{align*}
E^O=&\lim _{\ell\to +\infty}E^O(\ell),\\
\end{align*}
where
\begin{align*}
E^O(\ell):=&\int_{-\ell}^\ell dX\langle 0;M| T^O_{TT}(\ell) |0;M\rangle
=\sum_{n=1}^\infty{2\omega_n\over e^{\pi L\omega_n}-1}
=\dfrac{\ell}{3\pi L^2}(1+O(L/\ell))
\end{align*}
has been defined.

On the other hand, the von Neumann entropy $S^O(\rho^O)$ of the Gibbs state $\rho^O$ can be
formally computed as
\begin{align*}
  S^O(\rho^O)&=-{\rm Tr}^O(\rho^O\log\rho^O)
=\pi L   E^O 
+\log Z\\
&=\delta(0)\left[{\pi\over 3L}+\int_0^\infty d\omega\log(1-e^{-\pi L\omega})^{-2}\right]\\
&={2\pi\over 3L}\delta(0),
\end{align*}
where the trace is taken over the Fock space of the creation and annihilation operators 
$(b^{\pm \dagger}_{\omega},b^{\pm}_{\omega})$.
This may be called as the entanglement entropy of the diamond region $O$.
The divergent factor $\delta(0)$ here also arises in the limit 
\begin{align*}
  S^O(\rho^O)=\lim_{\ell\to +\infty}S^O(\rho^O;\ell),
\end{align*}
where
\begin{align*}
  S^O(\rho^O;\ell)=&\pi L E^O(\ell)+\log Z(\ell)
:=\pi L E^O(\ell)+ \sum_{n=1}^\infty \log(1-e^{-\pi L\omega_n})^{-2}\\
=&\dfrac{2\ell}{3L}(1+O(L/\ell)).
\end{align*}

In general, the amount of the entropy to energy ratio $S/E$ 
contained within a given finite region is believed to 
be bounded from above by 
the typical length scale $R$ of the region
as
\begin{align*}
  S/E<2\pi R.
\end{align*}
This is known as the Bekenstein bound~\cite{Be81}.
In the present case, we find the relationship
\begin{align*}
  S^O(\rho^O)=2\pi L E^O, 
\end{align*}
(in the sense that $S^O(\rho^O;\ell)=2\pi LE^O(\ell)(1+O(L/\ell))$ holds,)
among the entropy $S^O(\rho^O)$, the length scale $L$ and the energy
 $E^O$.
This shows that the present system saturates the Bekenstein bound.

\section{Final Remarks}
Finally, let us try to speculate on another interpretation
of the entropy $S^O(\rho^O)$ in terms of the information theory.
The trajectory of the modular flow is 
the curve: $X={\rm const.}$, which corresponds to the uniformly accelerated motion
with the proper acceleration $a=-L^{-1}\sinh(2X/L)$.
In other words, each congruence class of trajectories of the modular flow
under the action of the proper Poincar\'e group
is represented by the pair of parameters $(L,X)$.
Each trajectory of the modular flow 
defines a nonnegative function
\begin{align*}
  P(L,X;T)=\frac{1}{2L}\frac{du^{+}(T)}{dT}=\frac{1}{2L\cosh^2\left({X+T\over L}\right)}
\end{align*}
of the  modular parameter $T$ on the trajectory, which integrates to unity:
\begin{align*}
  \int_{-\infty}^{\infty}dT P(L,X;T)=1.
\end{align*}
We interpret this as determining a certain probability density associated with
the modular flow.
For example, if an observer $(L,X)$ following the modular flow regards the increase of
the Minkowski time $u^+$ as a probabilistic process, so that
$u^+$ jumps from $-L$ to $L$ once in his history at the modular time $T$,
he could expect that this jump occurs with the probability density $P(L,X;T)$.

Given the family of probability density functions $P(L,X;T)$,
the parameter space $(L,X)$ inherits the structure of the Riemannian manifold.
The Riemannian metric on the parameter space is given by the Fisher information
metric
\begin{align*}
  G_{ij}(L,X)=-\int_{-\infty}^\infty dT P(L,X;T)
{\partial^2\over \partial y^i \partial y^j}\log P(L,X;T),
\end{align*}
where $y^i=(L,X)$ denotes the coordinates on the parameter space.
In the present case, the parameter space $(L,X)$ turns out to be
the Poincar\'e half plane. In fact, the Fisher information metric has the
form
\begin{align*}
  G={(1+2\zeta(2))dL^2+4dX^2\over 3L^2}.
\end{align*}
The distance in the parameter space determined by the Fisher information metric
gives an invariant measure of the difference between a pair of
probability density functions.
Applying this to the 
 distant pair: $P=(L,-\ell)$ and $Q=(L,\ell)$
in the diamond region $O$ of the fixed size $L$, we get
\begin{align*}
  {\rm Dist}(P,Q)=\int_{-\ell}^\ell{2\over \sqrt{3}L} dX
={4\ell \over \sqrt{3}L} 
=2\sqrt{3}S^O(\rho^O;\ell)(1+O(L/\ell)).
\end{align*}
Thus, this amount of information discrepancy is proportional to the entanglement entropy.

We can also compute explicitly the Shannon entropy $S^{\rm Sh}(L)$ as
\begin{align*}
  S^{\rm Sh}(L)=\int_{-\infty}^\infty dTP(L,X;T)\log {1\over P(L,X;T)}=\log {e^2L\over 2},
\end{align*}
which is a function of $L$. This quantity can be also related with the
distance between the point $P=(L,X)$ and $P'=(L+\delta L,X)$ as
\begin{align*}
 {\rm  Dist}(P,P')&=\int_L ^{L+|\delta L|} \sqrt{1+2\zeta(2)\over 3}{dL\over  L}\\
&=\sqrt{1+2\zeta(2)\over 3}
[S^{\rm Sh}(L+|\delta L|)-S^{\rm Sh}(L)].
\end{align*}
Thus, the Shannon entropy is relevant to  the entropy correction~\cite{SU94,CW94} associated
with the variation of the size of the diamond region $O$.

In this way, the Riemannian structure of the parameter space 
of a certain kind of the probability density functions
might have to do with
the entanglement entropy of the subsystem and its corrections.
We hope this viewpoint provides  some insight into the better understanding 
of the information-theoretic origin of the Bekenstein-Hawking entropy of black holes.



\begin{thebibliography}{99}
\bibitem{DSW86}
W.~Driessler, S.~J.~Summers and E.~H.~Wichmann,
Commun.~Math.~Phys.~{\bf 105}, 49 (1986).

\bibitem{Ha96}
R.~Haag, 
{\it Local Quantum Physics: Fields, Particles, Algebras}
(Springer, Berlin, 1996).


\bibitem{MR03}
P.~Martinetti and C.~Rovelli,
Classical~Quantum~Gravity~{\bf 20} 4919 (2003).

\bibitem{Unruh}
S.~A.~Fulling,
Phys.~Rev.~D {\bf 7}, 2850 (1973);
P.~C.~W.~Davies,
J.~Phys.~A {\bf 8}, 609 (1975);
W.~G.~Unruh,
Phys.~Rev.~D {\bf 14}, 870 (1976).



\bibitem{BW75}
J.~J.~Bisognano and E.~H.~Wichmann,
J.~Math.~Phys. {\bf 16}, 985 (1975).

\bibitem{Tak70}
M.~Takesaki,
{\it Tomita's Theory of Modular Hilbert Algebras and its Applications},
Lecture Notes in Mathematics {\bf 128} (Springer, Berlin, 1970).



\bibitem{SW64}
R.~F.~Streater and A.~S.~Wightman,
{\it PCT, Spin and Statistics, and All That}
(W.~A.~Benjamin, Inc., New York, 1964).


\bibitem{HL82}
P.~D.~Hislop and R.~Longo,
Commun.~Math.~Phys.~{\bf 84}, 71 (1982).






\bibitem{BD84}
See e.g., N.~D.~Birrell and P.~C.~W.~Davies,
{\it Quantum Fields in Curved Space}
(Cambridge Univ. Press, Cambridge, 1984).


\bibitem{PW00}
M.~K.~Parikh and F.~Wilczek, 
Phys.~Rev.~Letters {\bf 85}, 5042 (2000).

\bibitem{BM09}
R.~Banerjee and B.~R.~Majhi,
Phys.~Lett.~B {\bf 675}, 243 (2009).


\bibitem{Be81}
J.~D.~Bekenstein,
Phys.~Rev.~D {\bf 23}, 287 (1981).

\bibitem{SU94}
L.~Susskind and J.~Uglum, 
Phys.~Rev.~D~{\bf 50},~2700 (1994).

\bibitem{CW94}
C.~Callan and F.~Wilczek,
Phys.~Lett.~B~{\bf 333},~55 (1994).

\end{thebibliography}
\end{document}